\documentclass[epj]{svjour}
\usepackage{graphics}
\begin{document}
\title{$\tau \to f_1 (1285)\pi^{-} \nu_{\tau}$ decay in the extended Nambu -- Jona-Lasinio model}
\author{M. K. Volkov 
             \and A. A. Pivovarov 
             \and A. A Osipov
}                     
%
%
\institute{Bogoliubov Laboratory of Theoretical Physics, Joint Institute for Nuclear Research, Dubna, 141980, Russia } 
\date{30 January 2018 / Revised version: date}
%
\abstract{
Within the framework of the extended Nambu -- Jona-Lasinio model, we calculate the matrix element of the $\tau \to f_1(1285) \pi^{-} \nu_{\tau}$ decay, obtain the invariant mass distribution of the $f_1\pi$ -system and estimate the branching ratio Br$(\tau \to f_1 \pi^{-}\nu_{\tau})=4.0\times 10^{-4}$. The two types of contributions are considered: the contact interaction, and the axial-vector $I^G(J^{PC})=1^-(1^{++})$ resonance exchange. The latter includes the ground $a_1(1260)$ state, and its first radially excited state, $a_1(1640)$. The corrections caused by the $\pi -a_1$ transitions are taken into account. Our estimate is in a good agreement with the latest empirical result Br$(\tau \to f_1 \pi^{-} \nu_{\tau})=(3.9\pm 0.5)\times 10^{-4}$. The distribution function obtained for the decay $\tau \to f_1(1285) \pi^{-} \nu_{\tau}$ shows a clear signal of $a_1(1640)$ resonance which should be compared with future experimental data including our estimate of the decay width $\Gamma (a_1(1640) \to f_1 \pi )=14.1\,\mbox{MeV}$.
} 
\maketitle
\section{Introduction}
\label{intro}
The recent measurements of the branching fractions of three-prong $\tau$ decay modes made by BABAR Collaboration \cite{Lees12} contain important new results on the decay $\tau \to f_1(1285) \pi^{-} \nu_{\tau}$ providing confidence that the precise information on the corresponding  invariant mass distributions will also soon become available. Together with the large data sets obtained on this mode in the past by BABAR \cite{Aubert08,Aubert05} and CLEO \cite{Bergfeld97} Collaborations this calls for improved understanding of the theoretical description of the process. The main purpose of our paper is to make a step in this direction.

The $\tau \to f_1\,\pi^{-} \nu_{\tau}$ decay is driven by the hadronization of the QCD axial-vector currents involved. The details of this mechanism are not yet clearly understood due to poor knowledge of the QCD dynamics at low-energies. Indeed, the invariant mass of the $(f_1, \pi )$-system belongs to the interval $m_{f_1}+m_\pi\leq \sqrt{s}\leq m_\tau$, so it is too low to apply the QCD perturbation theory, but it is too large that the original chiral perturbation theory (ChPT) of QCD would be applicable \cite{Weinberg79,Gasser84,Gasser85a,Gasser85b}. In addition, an order of magnitude of energies involved seems to indicate that not only the ground axial-vector $a_1(1260)$ state contributes to the pertinent hadronic axial-vector current. It is not excluded that the first radial excitation, $a_1(1640)$, of the ground resonance state may be also important. We have found that both resonances affect the form of the spectral function, giving an interesting interference picture.

Presently there is no clear understanding of the nature of the $a_1(1260)$ and $f_1(1285)$ mesons. In our work we hold the view that these resonances are the standard quark-antiquark bound states. Our reasoning is based on the large-$N_c$ expansion of QCD \cite{Hooft74,Witten79}, indicating that at $N_c=\infty$ mesons are pure $q\bar q$ states. Phenomenologically these states can be described by the local effective meson Lagrangians \cite{Schwinger67,Wess67,Weinberg68,Gasiorowicz69}. These Lagrangians are not known from first principles, however, they basically can be constructed on the chiral symmetry grounds. Comprehensive reviews of such attempts can be found in \cite{Meissner88,Bando88}. The general problem of including these states in ChPT has been addressed in \cite{Ecker89,Pich89}. Notice, that there is a different interpretation of $a_1(1260)$ and $f_1(1285)$. Assuming that these resonances not belong to the large-$N_c$ ground state of QCD (i.e. the $qq\bar q\bar q$ states), one can generate them in ChPT by implementing unitarity in coupled-channels (see, for instance,  \cite{Lutz04,Roca05,Zhou14,Xie15} and references therein). This approach can provide an alternative platform for studying of the $\tau\to f_1\pi^-\nu_\tau$ decay. 

Over last years, not much has been done with respect to the theoretical study of this particular mode of the $\tau$-decay. One can indicate just a few attempts. This is an approach based on a meson dominance model considered in \cite{Calderon13}, which leads to a strong disagreement with the data of BABAR Collaboration, namely the calculated branching ratio Br$(\tau \to f_1\, \pi^{-} \nu_{\tau})=1.3\times 10^{-4}$ is three times less of the experimental value Br$(\tau \to f_1\, \pi^{-} \nu_{\tau})=(3.9\pm 0.5)\times 10^{-4}$ reported by the Particle Data Group \cite{Patrignani16}. More encouraging but relatively old result, Br$(\tau \to f_1\, \pi^{-} \nu_{\tau})=2.91\times 10^{-4}$, gives the model \cite{Li97} based on the hypothesis of $a_1(1260)$-meson dominance.

Both ideas are naturally realized under the framework of the Nambu -- Jona-Lasinio (NJL) model \cite{Eguchi76,Volkov82,Ebert83,Volkov84,Volkov86,Ebert86,Vogl91,Klevansky92,Volkov93,Bijnens93,Hatsuda94,Volkov94,Osipov96}. The first attempt to apply this model to describe the hadron part of the $\tau \to f_1 (1285) \pi^{-} \nu_{\tau}$ decay is presented in \cite{Vishneva14}. Though an analysis made there allows to reproduce the experimental value of branching ratio, the picture presented cannot be considered as a fully satisfactory description. Here we improve it on the following aspects. 

First, as opposed to \cite{Vishneva14} we demonstrate that the pseudoscalar channel does not give contribution to the decay $\tau \to f_1 \pi^{-} \nu_{\tau}$. The latter is a direct consequence of the anomaly structure of the corresponding quark triangle diagram. Indeed, for the pseudoscalar-exchange channel the hadron axial-vector current, $J_\mu^A$, is proportional to a gradient of the pion field: $J_\mu^A\propto\partial_\mu\pi$. If the pion turns into the $f_1\pi$ couple, the corresponding $f_1\pi\pi$ vertex vanishes because it has an anomaly structure $e_{\mu\nu\alpha\beta}\epsilon^\mu_{f_1} p_1^\nu p_2^\alpha p_3^\beta$, where the antisymmetric Levi-Civita symbol is contracted with a polarization vector of the axial-vector field, $\epsilon^\mu_{f_1}$, and with three momenta $p_1, p_2, p_3$ of particles involved. However, due to the conservation of the total momenta, there are only two independent vectors. The linear dependence between $p_1, p_2$ and $p_3$ makes the product to be zero. On the other hand, if pion turns into $a_1$ (the $\pi-a_1$ transition is described by the Lagrangian density $\propto \partial_\nu \vec\pi\vec a_1^\nu$ \cite{Osipov17je,Osipov17ap,Morais17}), the amplitude of the $a_1\to f_1\pi$ transition, which is not zero by itself $e_{\mu\nu\alpha\beta}\epsilon^\mu_{f_1} \epsilon^\nu_{a_1} p_1^\alpha p_2^\beta\neq 0$, will be changed to $\epsilon^\nu_{a_1}\to p^\nu_{\pi}$ and this vanishes the product for the reasons just mentioned above.  

Second, we show that it is necessary to take into account the first radially-excited state of the $a_1(1260)$-meson, that is the $a_1(1640)$ resonance. This state has not been considered in \cite{Vishneva14}. The evidence of this hadron resonance has been recently approved by the new data of COMPASS collaboration \cite{Wallner17}. To take the $a_1(1640)$ into account, we carry out the calculations in the framework of the extended NJL model \cite{VolkovW97,Volkov97,VolkovE97,VolkovYu00,Volkov06,Volkov16,Volkov17}. The model allows one to describe both the ground and the first radially-excited meson states in accord with the chiral symmetry requirements. 

And finally, we study the influence of the $\pi -a_1$ transitions on the amplitude and demonstrate their significance. Notice the essential difference between present calculations of $\pi -a_1$ effects, and the previous ones, presented in \cite{Vishneva14}. In our work we take into account the $\pi -a_1$ transition on the external pion line, that has not been done in \cite{Vishneva14}. On the contrary, we show that the $\pi -a_1$ transition on the virtual axial-vector line, considered in that paper, does not contribute. Thus, we properly account for $\pi -a_1$ mixing effects in the $\tau \to f_1 \pi^{-} \nu_{\tau}$ decay amplitude.  
 
As a result we obtain a reasonable theoretical description of the branching ratio of the $\tau \to f_1\, \pi^{-} \nu_{\tau}$ decay. We believe that as soon as new experimental data on the spectral functions of this process will be available a more detailed test of the extended NJL model will be possible. 

The material of this paper is distributed as follows. In Section 2 we establish some convenient notations and review the basic properties of the extended NJL model including its Lagrangian, which is the basis of all our calculations. In Section 3 we give a derivation of the $\tau \to f_1 \pi^{-} \nu_{\tau}$ decay amplitude which does not take into account the $\pi -a_1$ mixing effects. To trace the numerical effect coming out of $\pi-a_1$ transitions, we intentionally delayed this material up to the Section 4, where we also calculate the spectral distribution of $f_1\pi$ pair and find  the two particle decay width of $a_1(1640)\to f_1(1285)\pi$.  In Section 5 we summarize our results and make conclusions. In the Appendix we present the integral form of the amplitude describing the $\tau \to f_1 \pi^{-} \nu_{\tau}$ decay and collect some general expressions for the quark-loop-integrals considered.   

\section{The quark-meson Lagrangian of the extended NJL model}
\label{sec:1}
Let us review the main ingredients of the model which we apply to study the  $\tau \to f_1\, \pi^{-} \nu_{\tau}$ decay. Its dynamics is described by the quark Lagrangian density ${\cal L}(x)$ with the effective $U(2)_L\times U(2)_R$ chiral symmetric four-quark interactions
\begin{eqnarray}
\label{qint}
{\cal L}(x)&=&\bar q(x) \left(i\hat\partial-m^0 \right)q(x)+{\cal L}_{int}(x), \nonumber \\
{\cal L}_{int}(x)&=&\frac{G_S}{2}\sum_{i=1}^{2}\sum_{a=0}^{3}\left\{[j^S_{a(i)}(x)]^2+[j^P_{a(i)}(x)]^2\right\} \nonumber \\
&-&\frac{G_V}{2}\sum_{i=1}^{2}\sum_{a=0}^{3}\left\{[j^V_{a(i)}(x)]^2+[j^A_{a(i)}(x)]^2\right\},
\end{eqnarray}  
where $q=(u,d)$ are the up and down current quark fields of mass $m^0=(m^0_u, m^0_d)$; $G_S$ is a coupling determining the strength of the four-quark interactions of scalar and pseudoscalar types, $G_V$ is a coupling of vector and axial-vector interactions. The general form of quark currents is
\begin{equation}
j^{S,P,V,A}_{a(i)}(x)=\!\!\int\! d^4x_1 d^4 x_2 \bar q(x_1) F^{S,P,V,A}_{a(i)}(x; x_1,x_2)q(x_2). 
\end{equation}
At each value of index $i$ the sum over $a$ in (\ref{qint}) is invariant with respect to chiral transformations. Thus, there are four independent $U(2)_L\times U(2)_R$ invariant interactions. 

The local interactions with $i=1$ represent the conventional NJL type model (see, for instance, \cite{Volkov86,Bijnens93}) which describes the physics of ground-state mesons with quantum numbers $J^{PC}=0^{++}, 0^{-+}, 1^{--}$ and $1^{++}$. The covariant form factors of local interactions are given by
\begin{eqnarray}
&&F^{S,P,V,A}_{a(1)}(x; x_1,x_2)=F^{S,P,V,A}_{a} \delta(x-x_1)\delta(x-x_2), \nonumber \\
&&F^{S,P,V,A}_{a} =\tau_a\, (1,i\gamma_5,\gamma_\mu, \gamma_\mu\gamma_5 ),
\end{eqnarray} 
where the flavour matrices $\tau_a=(1, \vec\tau)$, with the standard notation of the isospin Pauli matrices $\vec\tau$. 

The non-local part of the Lagrangian density $i=2$ represents the first radial excitations of ground-states. The corresponding covariant non-local form factors have been constructed in \cite{VolkovW97} with the use of the following transformation  
\begin{eqnarray}
F^{S,\ldots}_{a(2)}(x; x_1,x_2)&=&\!\int\!\frac{d^4k}{(2\pi )^4}\frac{d^4p}{(2\pi )^4} \exp i\left[ p\left(x-\frac{x_1+x_2}{2}\right) \right.\nonumber \\
&-& \left. k\left(x_1-x_2\right)\right] F^{S,\ldots}_{a(2)}(k_\perp ),
\end{eqnarray}
where $k_\mu$ is a relative momentum of the quark-antiquark pair, and $p_\mu$ is a 4-momentum of their center of mass reference frame, i.e. a meson momentum. The total momentum $p_\mu$ of the composite hadron provides a naturally preferred direction which forms the basis for a covariant three dimensional support to the interaction kernel $F^{S,\ldots}_{a(2)}(k_\perp )$ which is implemented here in accord with the concept of bilocal fields. The dependence of the interaction kernel on the transverse part of the quark momentum $k$, i.e. $k_\perp =k-(kp)p/p^2$ is a consequence of a subsidiary condition \cite{Markov40,Yukawa50}, which as it was shown in \cite{Lukierski77} is equivalent to a 'gauge principle' and expresses the redundance of the longitudinal component of the relative momentum $k$ for the physical interaction between the quark-antiquark constituents. 

The covariant kernels of non-local interactions are       
\begin{equation}
\label{ff2}
F^{S,P,V,A}_{a(2)}(k_\perp^2)=F^{S,P,V,A}_{a} c^{S,P,V,A} f (k_\perp^2 )\theta (\Lambda_3-|k_\perp |).
\end{equation} 
The coefficients $c^{S,P,V,A}$ renormalize the couplings of four-quark interactions $G_S$ and $G_V$ increasing a strength of these forces for the non-local interactions. They are fixed from the empirical values of meson masses. The step function $\theta (\Lambda_3-|k_\perp |)$, where $\Lambda_3$ is a covariant cutoff, restricts the integration over relative momentum of a bound quark-antiquark pair to the size of the bag. The function 
\begin{equation}
\label{fk}
f (k_\perp^2) =1+d |k_\perp |^2, \quad |k_\perp |=\sqrt{-k_\perp^2},
\end{equation} 
being a Lorentz scalar, can be calculated in any convenient reference frame. In the following we use the instantaneous rest frame of the meson. In this case $k_\perp =(0, \vec k)$, and $f (k_\perp^2)= 1+d \vec k^2\equiv f(\vec k^2)$. The slope parameter $d$ is fixed by the requirement that the numerical values of the quark condensate and constituent quark masses are not changed due to an inclusion of the radially excited states. Equivalently, one can require that the single excited quark-antiquark states averaged over vacuum are vanishing (an absence of the vacuum tadpoles). It gives $d = -1.784 \,\textrm{GeV}^{-2}$. There is a simple argument in favour of this requirement: the non-local bound states do not survive in the large $N_c$ limit, therefore they cannot affect the main characteristics of the QCD ground state. The form factor $f (k_\perp^2)$ has for $d\leq \Lambda_3^{-2}$ the form of an excited-state wave function, with a node in the interval $0\leq |k_\perp |\leq \Lambda_3$. In (\ref{fk}) we consider only the first two terms in a series of polynomials in $|k_\perp |^2$; inclusion of higher excited states would require polynomials of higher degree. 

The Lagrangian density (\ref{qint}) has to be bosonized. The boson variables can be introduced in the two stages. On the first stage, the four quark interactions can be equivalently rewritten as a Yukawa type quark-antiquark-meson interactions. In this form the model Lagrangian has a structure of a linear sigma model. On the second stage, one should integrate out the quark fields completely. Exactly this way the Lagrangian of the extended NJL model has been worked out in \cite{Volkov97,VolkovYu00}. Its numerous applications were reviewed in \cite{Volkov17}. Let us stress the main features of such calculations: (a) The model reveals the mechanism of spontaneous chiral symmetry breaking. Starting from some critical value of coupling $G_S\geq G_{crit}$ the Wigner-Weyl ground state of the system is changed to the Nambu-Goldstone phase. The transition is described by the gap equation. In particular, the current quark mass $m^0$ is replaced on the constituent quark mass $m$. It is assumed that the non-local sector of the model does not contribute to $m$, and does not affect the value of the quark condensate. This assumption help us to fix a slope parameter $d$; (b) Several mixing effects take place at the level of free Lagrangian. First, as a consequence of the phase transition, the mixing between $J^{P}=0^-$ and $J^P=1^+$ states, the so-called $\pi -a_1$ transitions, occurs. Second, there are mixings between ground and excited states with the same quantum numbers; (c) The effective meson vertices and corresponding coupling constants follow from the one-quark-loop calculations. It means that one should separate divergences in a regularized form, and renormalize the meson fields.  All these effects are taken into account in our calculations.     

The description of collective bound states can be facilitated if we introduce, as it was discussed above, the set of bosonic variables and pass to the semi-bosonized effective meson action which is responsible for the  $\tau \to f_1\, \pi^{-} \nu_{\tau}$ decay, namely, the part which describes both the interactions of  the ground pseudoscalar $\pi^{\pm}$, axial-vector $a_1^{\pm}(1260)$,  $f_1(1285)$ mesons and their first radially excited states with light constituent quarks. In momentum representation the action takes the form 
\begin{eqnarray}
\label{Lagrangiane}
S &=&\!\int\!\frac{d^4p}{(2\pi )^4}\bar q(p) \left(\hat p-m \right)q(p)+ \Delta S_{mass}+ \Delta S_{int}, \nonumber \\
\Delta S_{int} &=&\! \int\!\frac{d^4p}{(2\pi )^4}\!\int\!\frac{d^4k}{(2\pi )^4}\, \bar{q}\left(k+\frac{p}{2}\right) \nonumber \\
&\times& \left[ \frac{1}{2} \gamma^{\mu}\gamma^{5}\left(A_{f_{1}}f_{1\mu}(p) + B_{f_{1}}f^{'}_{1\mu}(p)\right) \right.\nonumber \\
&+& \frac{1}{2} \gamma^{\mu}\gamma^{5} \vec\tau \left(A_{a_{1}}\vec a_{1\mu}(p) + B_{a_{1}}\vec a'_{1\mu}(p)\right) \nonumber \\
&+&  \left. \gamma^{5} \vec\tau \left(A_{\pi}\vec \pi (p) + B_{\pi}\vec \pi'\right)\right]q\left(k-\frac{p}{2}\right) .
\end{eqnarray}
We shall not discuss here the mass part of the action $\Delta S_{mass}$ (see, for instance, \cite{Volkov17} or references therein). Our notations are as follows: $f_{1\mu}$, $\vec a_{1\mu}$ and $\vec \pi $ are the fields corresponding to the axial-vector and pseudoscalar mesons, the related excited states are marked with a prime. The constituent quark fields, $q$, have equal masses $m_{u} = m_{d} = 280$ MeV \cite{VolkovYu00,Volkov17}. The summation over colour index is assumed. The couplings of the ground state meson $M=(\pi, a_1, f_1)$ with quarks, $A_M$, and the corresponding couplings of its radially excited state $M'=(\pi', a_1', f_1')$,  $B_M$, can be written in a form 
\begin{eqnarray}
\label{coefficients}
&&\!\!\!\! A_M = A_M^0\left[g_M \sin\theta_{M}^+ +g'_M f(k_\perp^2)\sin\theta_M^ -\right]\! , \nonumber \\
&&\!\!\!\! B_M =-A_M^0\left[g_M \cos\theta_{M}^+ +g'_{M} f (k_\perp^2)\cos\theta_M^ -\right].
\end{eqnarray}
These expressions are obtained as a result of several procedures. The renormalization factors $g_M$ and $g_M'$ eliminate the divergent parts of the amplitudes describing the one-quark-loop self-energy transitions $M\to M$ and $M'\to M'$ correspondingly. In the NJL model they are fully determined by the requirement that such transitions generate a free Lagrangian of meson fields. Their specific values are expressed through the divergent quark-loop integrals
\begin{equation}
I_{2, n} =
-i\frac{N_{c}}{(2\pi)^{4}}\int\frac{f^{n}(k_\perp^2)}{(m^{2} - k^2)^{2}}\theta(\Lambda_{3}^2 - k_\perp^2)
\mathrm{d}^{4}k,
\end{equation}
which are regularized by imposing the three-dimensional cutoff on $|k_\perp|\leq \Lambda_3 = 1.03$ GeV (in accord with eq.(\ref{ff2})), after carrying out the $k_0$ integration in the self-energy diagrams \cite{Volkov97,Volkov17}. In particular, we have
\begin{eqnarray}
&&g_{f_1} = g_{a_{1}}=\sqrt{\frac{3}{2I_{2,0}}}, \quad g_{\pi}=\sqrt{\frac{Z_\pi}{4I_{2,0}}}, \nonumber\\
&&g'_{f_1} = g'_{a_1} = \sqrt{\frac{3}{2I_{2,2}}}, \quad g'_{\pi}=\sqrt{\frac{1}{4I_{2,2}}}.
\end{eqnarray}
Note that $Z_{\pi}$ is the factor induced by a diagonalization of the free Lagrangian for ground state mesons (the $\pi -a_{1}$ transitions). To avoid this mixing the unphysical axial-vector fields $\vec a_{1\mu}(x)$ should be redefined
\begin{equation}
\label{pa-trans}
\vec a_{1\mu}(x)=\vec a_{1\mu}^{phys}(x)+\sqrt{\frac{2Z_\pi}{3}}\kappa m\partial_\mu\vec\pi (x),
\end{equation}   
where 
\begin{eqnarray}
Z_{\pi} = \left(1 - \frac{6m^{2}}{M^{2}_{a_{1}}}\right)^{-1} \approx 1.45. 
\end{eqnarray}
A dimensional parameter $\kappa$ is fixed by requiring that the free meson Lagrangian does not contain the unphysical $\vec\pi-\vec a_{1\mu}$ transitions; it gives $\kappa =3/M_{a_1}^2$, where $M_{a_1}$ is a mass of the $a_1(1260)$ meson, $M_{a_1}=1230\pm 40\,\mbox{MeV}$. The replacement (\ref{pa-trans}) does not affect the Green function of the axial-vector field, but it affects the kinetic term of a pion free Lagrangian. Consequently, the pion field wave-function is additionally renormalized by the factor $\sqrt{Z_\pi}$ in (\ref{pa-trans}). 

There are other mixings between $J^P=1^+$ and $J^P=0^-$ states. The $\pi' -a_1'$ mixing does not contribute to the $\tau\to f_1\pi\nu_\tau$ decay. In the following we neglect the $a_1'\to\pi$ mixing due to a heavy mass of the $a_1'$ state which is associated with $a_1(1640)$ meson. 
 
Eq. (\ref{coefficients}) includes the angles, $\theta_{M}^{0}$ and $\theta_{M}$ \cite{Volkov97,Volkov17}. These parameters appear due to  $M\to M'$ mixing ($\theta_{M}^{0}$), and as a result of diagonalization, which aimed to avoid such a mixing ($\theta_{M}$). We arranged them in the following combinations: $A_M^0=1/\sin(2\theta_{M}^{0})$, $\theta_M^\pm = \theta_{M} \pm \theta_{M}^{0}$. The numerical values of mixing angles follow from the meson mass formulae. They are 
\begin{eqnarray}
\label{angles}
&& \theta_{f_{1}} = \theta_{a_{1}} = 81.8^{\circ}, \quad \theta_{\pi} = 59.48^{\circ},  \nonumber\\
&& \theta_{f_{1}}^{0} = \theta_{a_{1}}^{0} = 61.5^{\circ}, \quad \theta_{\pi}^{0} = 59.12^{\circ}.  
\end{eqnarray}
Notice, that all parameters of the model are determined from the empirical data, which are not related with the  characteristics of the processes considered in this work. Therefore, our estimations can be used to test the predictive power of the extended NJL model.    

\section{The decay amplitude without $\pi -a_{1}$ transitions}
\label{sec:2}
\begin{figure}
\resizebox{0.40\textwidth}{!}{%
  \includegraphics{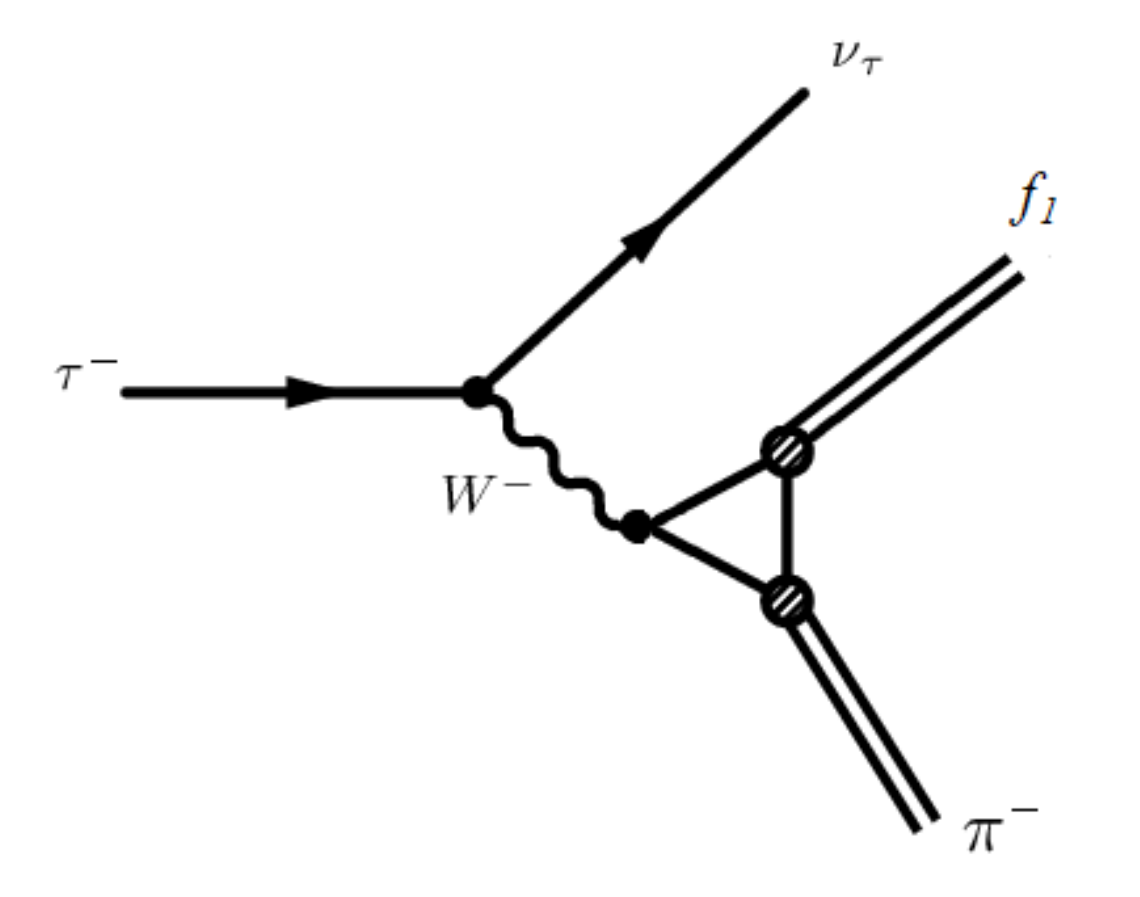}
}
\caption{The diagram which is responsible for the direct contribution to the amplitude of the $\tau \to f_1 \pi^{-} \nu_{\tau}$ decay.}
\label{Contact}      
\end{figure}

\begin{figure}
\resizebox{0.50\textwidth}{!}{%
  \includegraphics{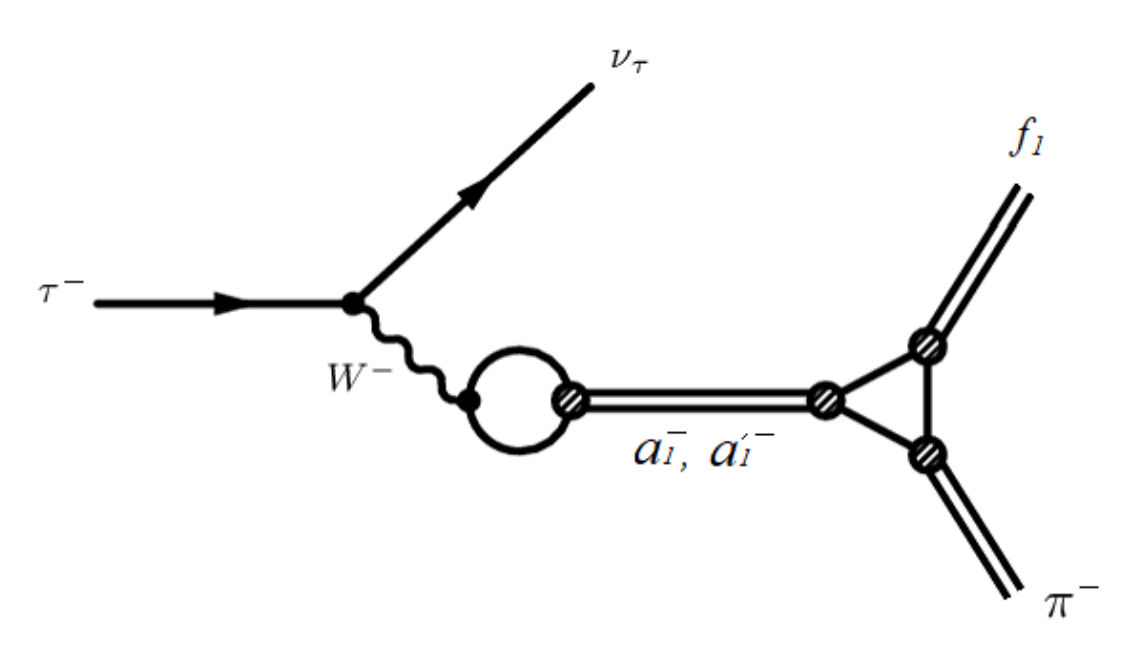}
}
\caption{The hadron resonance contributions to the $\tau \to f_1 \pi^{-} \nu_{\tau}$ decay. Here $a_1=a_1(1260)$ and $a_1'=a_1(1640)$ meson states.}
\label{Intermediate}      
\end{figure}

%
The decay amplitude of the process $\tau \to f_1(1285)\, \pi^- \nu_\tau$ is described by two types of diagrams which are shown in Fig.\ref{Contact}, and Fig.\ref{Intermediate} (the $\pi -a_1$ transitions are neglected there). The first diagram describes the so-called direct contribution to the amplitude. This means that $W$-boson decays directly to the final products of the reaction, $f_1\pi$ - pair, i.e. without a resonance exchange. The latter is taken into account by the second diagram. Notice, that hadrons are alway interact through the one-quark-loop in accord with our action (\ref{Lagrangiane}). In addition, we would like to point out that these loop integrals are expanded in momenta of external fields, and only the divergent parts are kept (in the case of anomalies, which lead to the finite result we take the first term of such expansion). This approximation is qualitatively justified by the $1/N_c$ expansion which states that meson physics in the large $N_c$ limit is described by the local vertices \cite{Hooft74,Witten79}. It is also well known from the sigma model that divergent parts of radiative corrections have a strictly chiral-symmetric structure \cite{Eguchi78}. 

The amplitude corresponding to the graphs shown in Fig.\ref{Contact} and Fig.\ref{Intermediate} is presented in an Appendix. Here we show the result obtained after derivative expansion of quark vertices, as it was discussed above. Note also that from (\ref{angles}) it follows that the angle $\theta_\pi^-$ is small, and $\theta_\pi\simeq \theta^0_\pi$. Neglecting $\theta_\pi^-$, one concludes from (\ref{coefficients}) that $A_{\pi} \simeq g_{\pi}$. As a result, an amplitude takes the form
\begin{eqnarray}
\label{M-without}
\mathcal{M} & = & 4mG_{F}V_{ud}\,g_\pi l^\mu e_{\mu\nu\alpha\beta}\epsilon^{\nu}(p_{f_1})p_{f_1}^\alpha p_{\pi}^\beta\left\{ I_{3}^{A_{f_{1}}} \right. \nonumber \\
&+& \frac{C_{a_1}}{g_{a_{1}}} I_{3}^{A_{f_{1}}A_{a_{1}}} \frac{s - 6m^2}{M_{a_{1}}^2 - s - i \sqrt{s}\Gamma_{a_{1}}} \nonumber\\
& + & \left. \frac{C_{a'_1}}{g_{a_{1}}} I_{3}^{A_{f_{1}}B_{a_{1}}} \frac{s - 6m^2}{M_{a'_{1}}^2 - s - i \sqrt{s}\Gamma_{a'_{1}}} \right\}\!.
\end{eqnarray}   
Here $G_F = 1.1663787(6) \times 10^{-11}$ MeV$^{-2}$ is the Fermi coupling constant; $V_{ud} = 0.97417\pm 0.00021$ is the Cabbibo-Kobayashi-Maskawa quark-mixing matrix element; $l_{\mu}=\bar\nu_\tau(Q')\gamma_\mu(1-\gamma_5)\tau (Q)$ is a lepton current, where $Q$ and $Q'$ are momenta of the tau-lepton and neutrino; $\epsilon_{\nu}(p_{f_{1}})$ is a polarization vector of the $f_1$ meson with the momentum $p_{f_1}$; $s = (p_{f_1} + p_{\pi})^2$ is a square of the invariant mass of the $f_1\pi$-pair,  the masses and widths of resonances are $M_{a_{1}} = 1230$ MeV, $M_{a'_{1}} = 1640$ MeV, $\Gamma_{a_{1}} \approx 400$ MeV, $\Gamma_{a'_{1}} = 254$ MeV \cite{Patrignani16}. The factors $C_{a_1}$ and $C_{a'_1}$ are the remaining of functions $A_{a_1}$ and $B_{a_1}$ (see eq.(\ref{coefficients})) after integration over an internal momentum in the $W-a_1$ and $W-a_1'$ quark-loops (see Fig.\ref{Intermediate}) correspondingly.
\begin{eqnarray}
C_{a_1} & = & A^0_{a_1} \left( \sin\theta_{a_1}^+ +R_V\sin\theta_{a_1}^- \right), \nonumber \\
C_{a'_1} & = & -A^0_{a_1} \left(\cos\theta_{a_1}^+ +R_V\cos\theta_{a_1}^-\right), \nonumber \\
R_{V} & = & \frac{I_{2,1}}{\sqrt{I_{2,0}I_{2,2}}}\, .
\end{eqnarray}
The integrals $I_3^{A_{f_1}\dots}$ come out from the evaluation of the one-loop-quark triangle diagrams of Fig.\ref{Contact} and Fig.\ref{Intermediate}. They correspond to the case $n=3$ of the general expression (in the next section the case $n=4$ will be also required)  
\begin{equation}
\label{In}
I_{n}^{A_{M}B_{M}\dots} =
\frac{-iN_c}{(2\pi)^4}\!\int\! \frac{A_M (k_\perp^2) B_M (k_\perp^2)\dots}{(m^2 - k^2)^n}\theta(\Lambda_3^2 - k_\perp^2)
\mathrm{d}^{4}k,
\end{equation}
where $A_{M}, B_{M}$ are given in (\ref{coefficients}). 

The amplitude (\ref{M-without}) has a conventional form $\mathcal{M}=G_{F}V_{ud}l^\mu H_\mu$, where we have found the hadron current $H_\mu$ with the help of the extended NJL model. The square of this amplitude has a simple form 
\begin{equation}
|\mathcal{M}|^2=4G_{F}^2V_{ud}^2\left[2(QH)(Q'H)-H^2 (QQ')\right].
\end{equation} 
Thus one can easily find the decay width of the process
\begin{equation}
\Gamma = \frac{1}{32M_\tau^3(2\pi )^3}\!\! \int\limits_{(M_{f_1}+M_\pi )^2}^{M_\tau^2}\!\!\!\!\! ds \int\limits_{t_-}^{t_+} dt \,| {\cal M} |^2
\end{equation}
by performing the numerical integration over kinematical variables $t=(Q-p_{f_1})^2$ and $s=(p_{f_1}+p_\pi )^2$.  Here a boundary of the physical region at fixed value of $s$ belongs to the interval $t_-\leq t\leq t_+$, where
\begin{eqnarray}
&&t_\pm = \frac{1}{2} \left( t_0\pm\sqrt{D}\right), \nonumber \\
&& t_0= M_\tau^2+M_{f_1}^2+M_\pi^2 -s -\frac{M_\tau^2}{s}\left(M_{f_1}^2-M_\pi^2\right), \nonumber \\
&&\sqrt{D}=\frac{1}{s}\left(M_\tau^2-s\right)\sqrt{\lambda (s,M_{f_1}^2,M_\pi^2 )}, \nonumber \\
&&\lambda (x, y, z )=[ x - (\surd y-\surd z )^2] [ x - (\surd y+ \surd z )^2 ]. 
\end{eqnarray}

Finally we arrive at the following result for the branching ratio Br$(\tau\to f_1\pi^-\nu_\tau )=6.04 \times 10^{-4}$. It can be schematically presented in terms of individual contributions as follows 
\begin{eqnarray}
&&10^{4}\times \mbox{Br}(\tau\to f_1\pi^-\nu_\tau)= 6.04 \nonumber\\
&&= |c|^2+|a_1|^2+|a'_1|^2+2\mbox{Re}(a_1 c^*+a'_1 c^*+a'_1a_1^*) \nonumber\\
&&= 2.25+8.43+0.78-7.34+0.53+1.39.
\end{eqnarray}
where $c, a_1, a'_1$ represent the contributions from the contact (direct) term, and from $a_1$ and $a'_1$ exchanges correspondingly. One can see that if one neglects the $\pi -a_1$ transitions (as we did here) the result overestimates the experimental value Br$(\tau \to f_1\, \pi^{-} \nu_{\tau})=(3.9\pm 0.5) \times 10^{-4}$.

\section{The effect of $\pi -a_{1}$ transitions}
\begin{figure}
\resizebox{0.30\textwidth}{!}{%
  \includegraphics{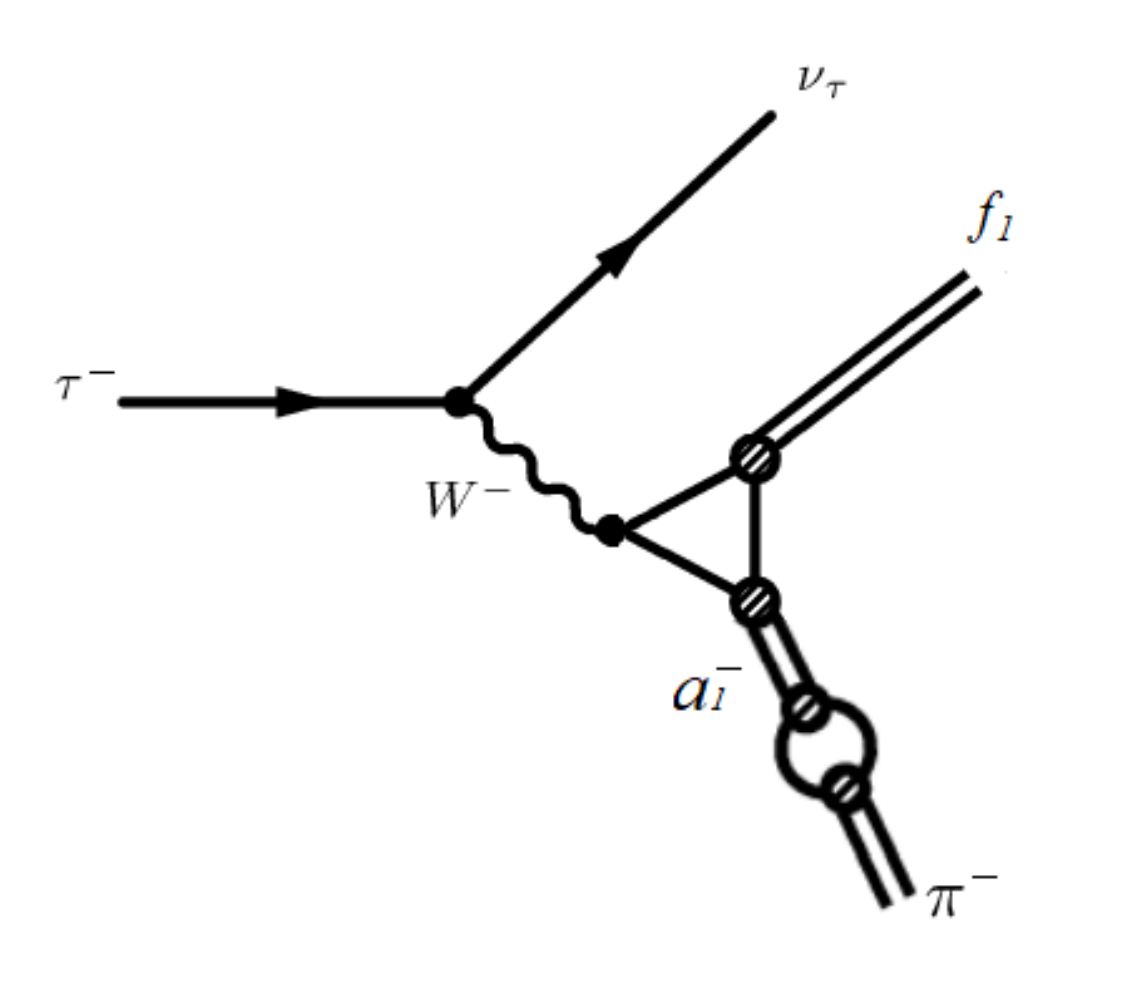}
}
\caption{The diagram which is responsible for the additional contribution to the direct term of the $\tau \to f_1 \pi^{-} \nu_{\tau}$ decay amplitude due to the $\pi a_1$ transitions.}
\label{pi-a1-contact}      
\end{figure}

The replacement (\ref{pa-trans}) in the quark-meson Lagrangian originates new vertices of the axial-vector type with the gradient of the pion field. This can be equivalently considered as a creation of the final pion via old $a_1$ meson field ($\pi -a_1$ transitions). To take them into account one has to append the diagrams shown in Fig.~\ref{pi-a1-contact} and Fig.~\ref{pi-a1-intermediate}. The corresponding additional contribution to the amplitude is given in the Appendix (see eq.(\ref{add-pi-a1})). It can be simplified after the following observation. 

\begin{figure}
\resizebox{0.40\textwidth}{!}{%
  \includegraphics{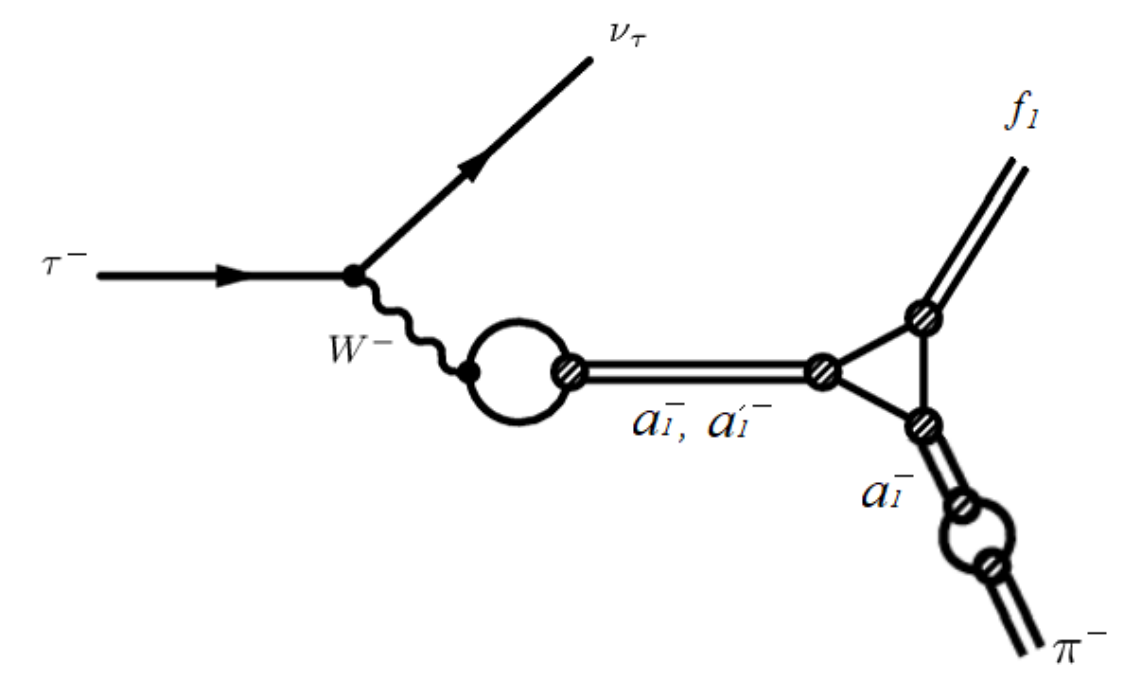}
}
\caption{The diagrams of the decay $\tau \to f_1 \pi^{-} \nu_{\tau}$ with the $a_1$ and $a'_1$ mesons exchange and $\pi -a_1$ transitions.}
\label{pi-a1-intermediate}      
\end{figure}

Let us consider the trace of the quark triangles corresponding to these diagrams  
\begin{equation}
\mbox{tr}[\gamma^{\nu}\gamma^{5}(\hat{k} + \hat{p}_{f_{1}} + m)\gamma^{\mu}\gamma^{5}(\hat{k} - \hat{p}_{\pi} + m) \gamma^{\lambda}\gamma^{5}(\hat{k} + m)].    
\end{equation}
It can be written as a sum
\begin{eqnarray}
\label{trace}
&& \mbox{tr}\,[\gamma^{\nu}(\hat{k} + \hat{p}_{f_{1}} + m)\gamma^{\mu}\gamma^{5}(\hat{k} - \hat{p}_{\pi} + m)\gamma^{\lambda}(\hat{k} + m) \nonumber\\
&& - 2m \gamma^{\nu}(\hat{k} + \hat{p}_{f_{1}} + m)\gamma^{\mu}\gamma^{5}(\hat{k} - \hat{p}_{\pi} + m)\gamma^{\lambda}].
\end{eqnarray}
It is not difficult to see that the first term coincides with a trace from the triangle of the process $f_{1}(1285) \to \rho \gamma$ considered in our previous work \cite{Osipov17}. After integration over loop-momentum $k$ in the corresponding quark triangle expression (see, for instance, eq.(\ref{afa})) one finds that it contributes to the amplitude as
\begin{eqnarray}
&&e^{\mu\nu\lambda}_{\, . \, . \ . \, \alpha} \, p^\alpha_{\pi} \left(2 p_{f_{1}}^{2} + p_\pi p_{f_{1}}\right) - e^{\mu\nu\lambda}_{\, . \, . \ . \, \alpha}\, p^\alpha_{f_{1}} \! \left( p_\pi p_{f_{1}} \right) \nonumber \\
&& + e^{\mu\nu}_{\, . \, .\, \alpha\beta}\, p^\alpha_{\pi} p^\beta_{f_{1}} p_{f_{1}}^{\lambda} - e^{\lambda\mu}_{\, . \, .\, \alpha\beta}\,   p^\alpha_{f_1} p^\beta_{\pi} p_{\pi}^{\nu}.
\end{eqnarray}
Due to $\pi -a_1$ transitions this result is multiplied by the 4-momentum $p_{\pi}^{\lambda}$. That vanishes it. 

The second term in eq. (\ref{trace}) is easily calculated, giving  $-8im^2e^{\mu\nu\lambda\alpha} (2k+p_{f_1}-p_\pi )_\alpha$. The result of its integration over $k$ depends on the structure of the integrand. In the case considered, we have three similar structures which differ only by the functions $A_M$ and $B_M$ at the vertices of anomalous triangle diagrams. These alternatives are absorbed by the corresponding coefficient $I_{4}^{A_{f_1}A_{\alpha_1}\dots}$ in the following common structure     
\begin{eqnarray}
8 m^{4} I_{4}^{A_{f_1}A_{\alpha_1}\dots} e^{\mu\nu\lambda}_{\, . \, . \ . \, \alpha}\, p^\alpha_{f_{1}},
\end{eqnarray}
where the integrals $I_{4}^{A_{f_1}A_{\alpha_1}\dots}$ are given by (\ref{In}).

Taking into account these remarks, one obtains the corrections induced by the $\pi -a_1$ transitions to the total amplitude of the $\tau \to f_1 \pi^{-} \nu_{\tau}$ decay. As a result the total amplitude is  
\begin{eqnarray}
\label{Mtot}
&& \mathcal{M}_{tot}  =  G_{F}V_{ud}l^\mu 4mg_\pi \left\{ \left[I_{3}^{A_{f_{1}}} - \frac{C_{a_1}6m^4}{g_{a_1}M_{a_1}^2} I_4^{A_{f_{1}}A_{a_{1}}} \right] \right. \nonumber\\
&&+ \frac{C_{a_1}}{g_{a_1}} \left[I_{3}^{A_{f_{1}}A_{a_{1}}} - \frac{C_{a_1}6m^4}{g_{a_1}M_{a_1}^2} I_{4}^{A_{f_{1}}A_{a_{1}}A_{a_{1}}}\right] \nonumber \\
&&\times \frac{s - 6m^2}{M_{a_{1}}^2 - s - i \sqrt{s}\Gamma_{a_{1}}} \nonumber\\
&&+  \frac{C_{a'_1}}{g_{a_{1}}} \left[I_{3}^{A_{f_{1}}B_{a_{1}}} - \frac{C_{a_1}6m^4}{g_{a_1}M_{a_1}^2} I_{4}^{A_{f_{1}}B_{a_{1}}A_{a_{1}}}\right]  \nonumber\\
&&\times\left.\frac{s - 6m^2}{M_{a'_{1}}^2 - s - i \sqrt{s}\Gamma_{a'_{1}}} \right\}  e_{\mu\nu\alpha\beta} \epsilon^{\nu}(p_{f_{1}}) p^\alpha_{f_{1} }p^\beta_{\pi}.
\end{eqnarray}
Here, in the first square brackets, the contact contribution is presented. The second and third square brackets contain the contributions of $a_1(1260)$ and $a_1(1640)$ resonances correspondingly.   

The numerical integration over the $ f_1\pi\nu_\tau$ three-body phase space with the amplitude (\ref{Mtot}) gives the branching ratio Br$(\tau\to f_1\pi^-\nu_\tau )=3.98 \times 10^{-4}$. The result can be schematically presented in terms of the individual contributions as follows 
\begin{eqnarray}
&&10^{4}\times \mbox{Br}(\tau\to f_1\pi^-\nu_\tau)= 3.98 \nonumber\\
&&= |c|^2+|a_1|^2+|a'_1|^2+2\mbox{Re}(a_1 c^*+a'_1 c^*+a'_1a_1^*) \nonumber\\
&&=1.62+5.92+0.45-5.22+0.33 +0.88,
\end{eqnarray}
where $c, a_1, a'_1$ represent the contributions from the contact term, and from $a_1$ and $a'_1$ exchanges correspondingly. One can see that if one takes into account the $\pi -a_1$ transitions the branching ratio is in a good agreement with the experimental value Br$(\tau \to f_1\, \pi^{-} \nu_{\tau})=(3.9\pm 0.5) \times 10^{-4}$.

\begin{figure}
\resizebox{0.50\textwidth}{!}{%
  \includegraphics{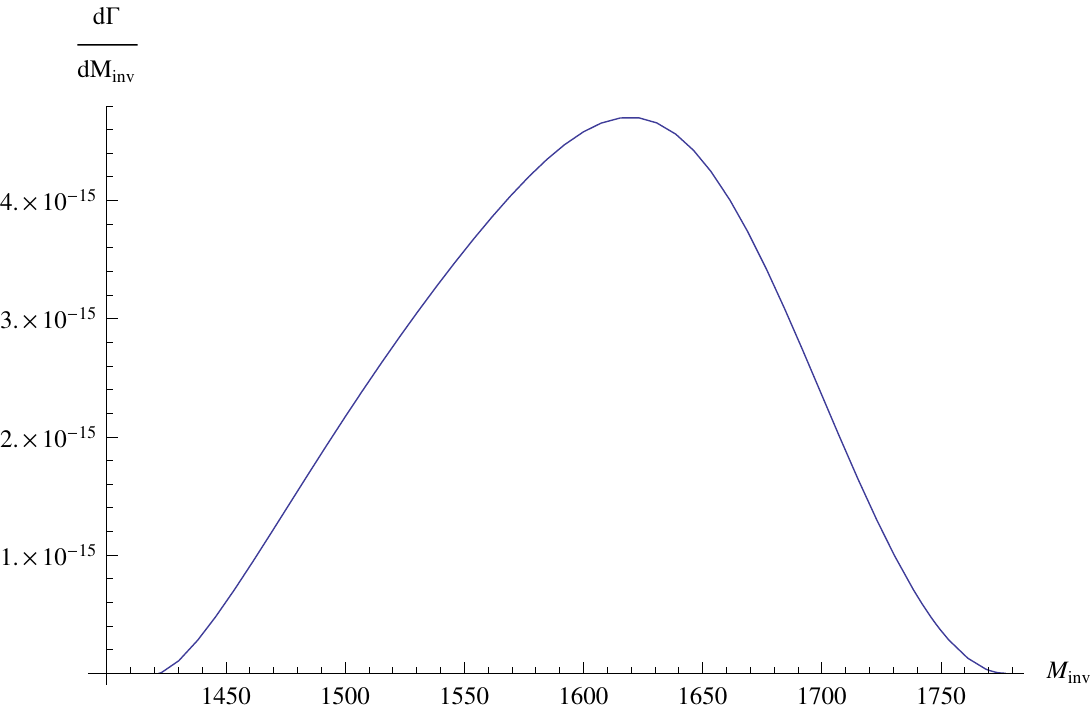}
}
\caption{The differential decay distribution of the $\tau \to f_1\pi^{-} \nu_{\tau}$ decay as a function of the invariant mass of $f_1\pi^-$- system, $M_{inv}=\sqrt{s}$.}
\label{distribution}      
\end{figure}

The differential decay distribution, $d\Gamma /d \sqrt s$, is shown in Fig. \ref{distribution}. It reaches its maximum value near the $a_1(1640)$ resonance mass shell. Therefore, it appears worth while to 
estimate the decay width of the state, that dominates the spectral function. The amplitude of the $a_1(1640)\to f_1(1285)\pi$ decay is given by 
\begin{eqnarray}
\mathcal{M} & = & 4img_\pi e_{\mu\nu\alpha\beta }\, \epsilon^{\mu}(p_{a_{1}})\epsilon^{\nu}(p_{f_{1}})  p^\alpha_{f_{1}}p^\beta_{\pi}  \nonumber \\
&\times& \left\{I_{3}^{A_{f_{1}}B_{a_{1}}} - \frac{C_{a_1}6m^4}{g_{a_{1} M_{a_1}^2}} I_{4}^{A_{f_{1}}B_{a_{1}}A_{a_{1}}} \right\}. 
\end{eqnarray}
It follows then that 
\begin{equation}
\Gamma \left[a_1(1640)\to f_1(1285)\pi \right]=14.1\,\mbox{MeV}.
\end{equation} 
The future measurements will show how good is the extended NJL model in its predictions here. Probably, the tau decay mode studied in this work can serve us with a detailed information on the nature of $a_1(1640)$ state.  

\section{Conclusions}
The main purpose of our calculations was to apply the extended NJL model to the decay $\tau\to f_1(1285)\,\pi^- \nu_\tau$. Presently available phenomenological data on this mode give a rare opportunity to test the model. As a result we have found that hadronic part of the amplitude is sensitive to the four types of different contributions which are equally important. These are the contact interaction, the exchange by the $a_1(1260)$ meson, the exchange by the first radially excited state, $a_1(1640)$, and the pion creation by the intermediate $a_1(1260)$ meson (the $\pi  a_1$-transitions). Indeed, the contact term alone gives the value Br$\ =2.25\times 10^{-4}$. The $a_1(1260)$ ground state exchange alone gives a larger number Br$\ =8.43\times 10^{-4}$. The sum of these two contributions is Br$\ =3.34\times 10^{-4}$. The radially excited state increase this value up to Br$\ =6.04\times 10^{-4}$. And finally taking into account the $\pi  -a_1$ transitions we come to the final number Br$\ =3.98\times 10^{-4}$ which agrees with presently known empirical values. 

Our result shows that both $a_1(1260)$ and $a_1(1640)$ exchanges are very important for the description of data, and it indicates that many ideas about description of the first radial exited meson states in the NJL model seems to be correct. Let us stress that we did not introduce any new parameters to get the consistent result. All model parameters were fixed with the use of other input data, which are not related with the process $\tau\to f_1(1285)\,\pi^- \nu_\tau$. Note, that this is not the only case where the model predictions correspond to the empirical values. We refer to the recent review \cite{Volkov17}, which contains many other examples. Those include the meson production processes in $e^+e^-$ collisions and tau decay modes.

Of some relevance may be the fact that our result does not leave the place for the contribution of a new axial-vector resonance $a_1(1420)$ observed recently by the COMPASS collaboration \cite{Adolph15}. Our study shows that even if there is a contribution due to $a_1(1420)$ exchange to the process discussed here, this contribution is negligible. It indicates that most probably $a_1(1420)$ is not the $q\bar q$ state. Some reasonings in favour of multi-quark structure of this state are recently given in \cite{Ivanov17}. 	

\section*{Acknowledgments}
We would like to thank A. B. Arbuzov for his interest to this work and useful discussions. 

\section*{Appendix}
Here we present the general expressions for the amplitude of $\tau\to f_1(1285)\pi^-\nu_\tau$ decay shown in Figs.\ref{Contact}-\ref{pi-a1-intermediate}. In the text we make derivative expansions of these expressions to obtain effective meson vertices in the long wavelength approximation. We consider the process 
\begin{equation}
\tau (Q)\to \nu_\tau (Q') f_1(p_{f_1}) \pi (p_\pi )
\end{equation} 
with the quantities in the parentheses denoting the four moments of the particles.   

The amplitude, corresponding to diagrams of Fig.\ref{Contact}-\ref{Intermediate}, has the following form
\begin{equation}
\mathcal{M} =  G_{F} V_{ud} l_{\mu} \mathcal{H}^\mu ,
\end{equation}
where a hadron current $\mathcal{H}^\mu$ is 
\begin{eqnarray}
\mathcal{H}^\mu & = & \frac{g_{\pi}}{2} \epsilon_{\nu}(p_{f_{1}}) 
\left\{ I_{Wf_{1}\pi 1}^{\mu\nu} + I_{Wf_{1}\pi 2}^{\mu\nu} \right. \nonumber\\
& + & \frac{I_{Wa_{1}}^{\mu\lambda}\left(I_{a_{1}f_{1}\pi1}^{\lambda\nu} + I_{a_{1}f_{1}\pi2}^{\lambda\nu}\right)}{M_{a_{1}}^2 - s - i \sqrt{s}\Gamma_{a_{1}}}  \nonumber \\
& + & \left.
 \frac{I_{Wa'_{1}}^{\mu\lambda}\left(I_{a'_{1}f_{1}\pi1}^{\lambda\nu} + I_{a'_{1}f_{1}\pi2}^{\lambda\nu}\right) }{M_{a'_{1}}^2 - s - i \sqrt{s}\Gamma_{a'_{1}}} 
\right\}.
\end{eqnarray}

The first two integrals $I_{W f_1\pi 1}^{\mu\nu}$ and $I_{Wf_{1}\pi 2}^{\mu\nu}$ describe the direct contribution from the transition $W^\mu\to f_1^\nu\pi$ generated by the triangle quark loop (see Fig.\ref{Contact}). In accord with two different directions for the loop momenta we specify them by indices 1 (clockwise) and 2 (counter-clockwise). Their expressions are 
\begin{eqnarray}
&& I_{Wf_{1}\pi1}^{\mu\nu} = N_c \int  \frac{\mathrm{d}^4k}{(2\pi)^4} A_{f_1}(k_\perp^2)  \\
&&\times\frac{ \mbox{tr}[\gamma^\nu\gamma^5(\hat{k} + \hat{p}_{f_{1}} + m)\gamma^{\mu}\gamma^{5}(\hat{k} - \hat{p}_{\pi} + m)\gamma^{5}(\hat{k} + m)]}{[(k + p_{f_{1}})^2 - m^2][(k - p_{\pi})^2 - m^2](k^2 - m^2)}, \nonumber 
\end{eqnarray}
\begin{eqnarray}
&&I_{Wf_{1}\pi2}^{\mu\nu} = N_c \int \frac{\mathrm{d}^4k}{(2\pi)^4} A_{f_{1}} (k_\perp^2) \\
&&\times\frac{\mbox{tr}[\gamma^{5}(\hat{k} + \hat{p}_{\pi} + m)\gamma^{\mu}\gamma^{5}(\hat{k} - \hat{p}_{f_{1}} + m)\gamma^{\nu}\gamma^{5}(\hat{k} + m)]}{[(k - p_{f_{1}})^2 - m^2][(k + p_{\pi})^2 - m^2](k^2 - m^2)} . \nonumber
\end{eqnarray}
 
The other two integrals $I_{Wa_1}^{\mu\lambda}$ and $I_{Wa'_1}^{\mu\lambda}$ describe the $W^\mu\to a_1^\lambda$ and $W^\mu\to {a'_1}^\lambda$ transitions correspondingly (see Fig.\ref{Intermediate}). The first of them is equal to
\begin{eqnarray}
\label{IWa}
&&I_{Wa_{1}}^{\mu\lambda} = N_c \int \frac{\mathrm{d}^4k}{(2\pi)^4}A_{a_{1}} (k_\perp^2) \nonumber \\
&&\times\frac{ \mbox{tr} [\gamma^{\lambda}\gamma^{5}(\hat{k} + \frac{\hat{q}}{2} + m)\gamma^{\mu}\gamma^5(\hat{k} - \frac{\hat{q}}{2} + m)]}{[(k + \frac{q}{2})^2 - m^2][(k - \frac{q}{2})^2 - m^2]}, 
\end{eqnarray}
with $q=Q-Q'=p_{f_1}+p_\pi$, and $q^2=s$. The second one can be obtained from (\ref{IWa}) by the replacement $A_{a_{1}} (k_\perp^2) \to B_{a_{1}} (k_\perp^2)$.

The anomalous quark triangle integrals $I_{a_{1}f_{1}\pi1}^{\lambda\nu}$ and $I_{a_{1}f_{1}\pi2}^{\lambda\nu}$ of Fig.\ref{Intermediate} differ from other similar pair $I_{a'_{1}f_{1}\pi1}^{\lambda\nu}$ and $I_{a'_{1}f_{1}\pi2}^{\lambda\nu}$ by the replacement $A_{a_{1}}(k_\perp^2)\to B_{a_{1}}(k_\perp^2)$. Therefore, we give here only the first two expressions  
\begin{eqnarray}
&& I_{a_{1}f_{1}\pi1}^{\lambda\nu} = N_c \int \frac{\mathrm{d}^4k}{(2\pi)^4}A_{a_{1}} (k_\perp^2) A_{f_{1}}(k_\perp^2) \\
&&\times\frac{\mbox{tr} [\gamma^{\nu}\gamma^{5}(\hat{k} + \hat{p}_{f_{1}} + m)\gamma^{\lambda}\gamma^{5}(\hat{k} - \hat{p}_{\pi} + m)\gamma^{5}(\hat{k} + m)]}{[(k + p_{f_{1}})^2 - m^2][(k - p_{\pi})^2 - m^2](k^2 - m^2)},\nonumber
\end{eqnarray}
\begin{eqnarray}
&&I_{a_{1}f_{1}\pi2}^{\lambda\nu} = N_c \int \frac{\mathrm{d}^4k}{(2\pi)^4} A_{a_{1}} (k_\perp^2) A_{f_{1}}(k_\perp^2) \\
&&\times\frac{ \mbox{tr}[\gamma^{5}(\hat{k} + \hat{p}_{\pi} + m)\gamma^{\lambda}\gamma^{5}(\hat{k} - \hat{p}_{f_{1}} + m)\gamma^{\nu}\gamma^{5}(\hat{k} + m)]}{[(k - p_{f_{1}})^2 - m^2][(k + p_{\pi})^2 - m^2](k^2 - m^2)}. \nonumber 
\end{eqnarray}

Consider now the diagrams shown in Fig.\ref{pi-a1-contact}-\ref{pi-a1-intermediate}. They give an additional contribution to the axial current $\mathcal{H}^\mu$ by taking into account the $\pi -a_1$ transitions
\begin{eqnarray}
\label{add-pi-a1}
\Delta\mathcal{H}^\mu &=& \frac{g_{\pi}}{4} \epsilon_{\nu}(p_{f_{1}})\frac{I_{a_{1}\pi}^{\lambda}}{M_{a_1}^2}\left\{
I_{Wf_{1}a_{1}1}^{\mu\nu\lambda} + I_{Wf_{1}a_{1}2}^{\mu\nu\lambda} 
\right. \nonumber\\
&+& \frac{I_{Wa_{1}}^{\mu\delta}\left(I_{a_{1}f_{1}a_{1}1}^{\delta\nu\lambda} + I_{a_{1}f_{1}a_{1}2}^{\delta\nu\lambda}\right)}{M_{a_{1}}^2 - s - i \sqrt{s}\Gamma_{a_{1}}}  \nonumber\\
&+&  \left.
\frac{I_{Wa'_{1}}^{\mu\delta} \left(I_{a'_{1}f_{1}a_{1}1}^{\delta\nu\lambda} + I_{a'_{1}f_{1}a_{1}2}^{\delta\nu\lambda}\right)}{M_{a'_{1}}^2 - s - i \sqrt{s}\Gamma_{a'_{1}}} \right\}
\end{eqnarray}
Here, the integrals with two Lorentz indices have been already defined in (\ref{IWa}). The one-index integral describes the quark loop corresponding to the $\pi -a_1$ transition, i.e.  
\begin{eqnarray}
&& I_{a_{1}\pi}^{\lambda} = N_c \int \frac{\mathrm{d}^4k}{(2\pi)^4} A_{a_{1}} (k_\perp^2) \nonumber \\
&&\times  \frac{ \mbox{tr} [\gamma^{5}(\hat{k} + \frac{\hat{p}_{\pi}}{2} + m)\gamma^{\lambda}\gamma^{5}(\hat{k} - \frac{\hat{p}_{\pi}}{2} + m)]}{[(k + \frac{p_{\pi}}{2})^2 - m^2][(k - \frac{p_{\pi}}{2})^2 - m^2]} . 
\end{eqnarray}

The integral $I_{Wf_{1}a_{1}2}^{\mu\nu\lambda}$ can be obtained from the integral
\begin{eqnarray}
&& I_{Wf_{1}a_{1}1}^{\mu\nu\lambda} = N_c \int  \frac{\mathrm{d}^4k}{(2\pi)^4} A_{f_{1}}(k_\perp^2)A_{a_{1}} (k_\perp^2)  \\
&&  \frac{ \mbox{tr}[\gamma^{\nu}\gamma^{5}(\hat{k} + \hat{p}_{f_{1}} + m)\gamma^{\mu}\gamma^{5}(\hat{k} - \hat{p}_{\pi} + m)\gamma^{\lambda}\gamma^{5}(\hat{k} + m)]}{(k^2 - m^2)[(k + p_{f_{1}})^2 - m^2][(k - p_{\pi})^2 - m^2]} \nonumber
\end{eqnarray}
by replacements $\nu\leftrightarrow \lambda$ and $p_{f_1} \leftrightarrow p_\pi$. Both describe the direct transition $W^\mu\to f_1^\nu a_1^\lambda$. 

The last four integrals can be obtained, for instance, from the integral 
\begin{eqnarray}
\label{afa}
&& I_{a_{1}f_{1}a_{1}1}^{\delta\nu\lambda} = N_c \int \frac{\mathrm{d}^4k}{(2\pi)^4} A_{a_{1}} (k_\perp^2) A_{a_{1}}(k_\perp^2) A_{f_{1}}(k_\perp^2) \\
&& \frac{\mbox{tr}[\gamma^{\nu}\gamma^{5}(\hat{k} + \hat{p}_{f_{1}} + m)\gamma^{\delta}\gamma^{5}(\hat{k} - \hat{p}_{\pi} + m)\gamma^{\lambda}\gamma^{5}(\hat{k} + m)]}{(k^2 - m^2)[(k + p_{f_{1}})^2 - m^2][(k - p_{\pi})^2 - m^2]}. \nonumber 
\end{eqnarray}
In this case $I_{a_{1}f_{1}a_{1}2}^{\delta\nu\lambda}$ is given by the replacement $\nu\leftrightarrow \lambda$ and $p_{f_1} \leftrightarrow p_\pi$; the integral $I_{a'_{1}f_{1}a_{1}1}^{\delta\nu\lambda}$ follows from (\ref{afa}) by the replacement of one of the two functions $A_{a_{1}} (k_\perp^2) $ by $B_{a_{1}} (k_\perp^2) $; and finally, the integral $I_{a'_{1}f_{1}a_{1}2}^{\delta\nu\lambda}$ can be obtained from (\ref{afa}) by the replacement of one of the two functions $A_{a_{1}} (k_\perp^2) $ by $B_{a_{1}} (k_\perp^2) $ together with $\nu\leftrightarrow \lambda$ and $p_{f_1} \leftrightarrow p_\pi$.
  
%
%

\end{document}